\documentclass[12pt,preprint]{aastex}

\usepackage{CJK}

\shorttitle{Moving group detection}
\shortauthors{Zhao et al.}

\begin{document}
\begin{CJK}{UTF8}{gbsn}

\title{Three Moving Groups Detected in the LAMOST DR1 Archive}

\author{J. K. Zhao \altaffilmark{1}, G. Zhao\altaffilmark{1}, Y. Q. Chen\altaffilmark{1}, T. D. Oswalt\altaffilmark{2}, K. F. Tan\altaffilmark{1}, Y. Zhang\altaffilmark{3}}


\altaffiltext{1}{Key Laboratory of Optical Astronomy, National Astronomical Observatories, Chinese Academy of Sciences,
Beijing 100012, China. zjk@nao.cas.cn}

\altaffiltext{2}{Embry-Riddle Aeronautical University 600 S. Clyde Morris Blvd. Daytona Beach FL, USA, 32114. oswaltt1@erau.edu}
\altaffiltext{3}{Nanjing Institute of Astronomical Optics $\&$ Technology, National Astronomical Observatories, Chinese Academy of Sciences, Nanjing 210042, China}

\begin{abstract}
We analyze the kinematics of thick disk and halo stars observed by the Large sky Area Multi-Object Fiber Spectroscopic Telescope. We have constructed a sample of 7,993 F, G and K nearby main-sequence stars (\textit{d} $<$ 2 kpc) with  estimates of position (x, y, z) and space velocity ($U$, $V$, $W$) based on color and proper motion from the SDSS DR9 catalog. Three `phase-space overdensities' are identified in [\textit{V}, $\sqrt{U^{2}+2V^{2}}$] with significance levels of $\sigma$ $>$ 3. 
  Two of them (Hyades-Pleiades stream, Arcturus-AF06 stream) have been identified previously. We also find evidence for a new stream (centered at \textit{V} $\sim$ -180 km s$^{-1}$) in the halo. The formation mechanisms of these three streams are analyzed. Our results support the hypothesis the Arcturus-AF06 stream and the new stream originated from the debris of a disrupted satellite, while Hyades-Pleiades stream has a dynamical origin. 
\end{abstract}


\keywords{Galaxy: kinematics and dynamics}



\section{Introduction}
Moving groups (MGs) or dynamic streams are stars that share the same velocity components but  have no spatial or overdensity center discernible from the field stars.  MGs are often called kinematic structures since the stars
that belong to these streams have the same motion. The concept of MGs originates from the work of Eggen (1996 and references therein). With the data from large surveys like Hipparcos (Perryman et al. 1997; van Leeuwen 2007), the Geneva-Copenhagen Survey (Nordstr\"{o}m et al. 2004), the Radial Velocity Experiment (RAVE; Steinmetz et al. 2006) and the Sloan Digital Sky Survey (SDSS; York et al. 2000), dozens of MGs have been confirmed.

Two main hypotheses have been proposed as the origin of MGs. MGs in the halo are generally attributed to the relics of satellite galaxies that have been disrupted by the Galaxy's tidal potential (Helmi et al. 1999). Those in the thin disk are thought to have been formed through dynamic interactions (Dehnen 1998), or from dissociated clusters (e.g., HR1614; De Silva 2007). Among MGs known to date, two are in thick disk:
 Arcturus stream (Navarro et al. 2004) and AF06 (Arifyanto $\&$ Fuchs 2006). However, the origin of these MGs are still disputed.  Navarro et al. (2004) reanalyzed the group of stars kinematically associated  with Arcturus stream and concluded that they constitute a peculiar grouping of metal-poor stars with similar apocentric radius, common angular momentum, and distinct metal abundance
patterns. Their results suggest the angular momentum of Arcturus stream is too low to arise from dynamical perturbations induced by the Galactic bar, suggesting a tidal origin for this MG. This hypothesis is supported by the well-defined sequence of abundance ratios of member stars taken from the available Gratton et al. (2003) catalog. Arifyanto $\&$ Fuchs (2006), who later detected the  Arcturus stream group in a compilation of various catalogs using their (\textit{V}, $\sqrt{U^{2}+2V^{2}}$) method, suggested an external origin as well, based on the goodness of theoretical 12 Gyr isochrone fits to the putative member stars. However, recent detailed abundance studies including various $\alpha$-process and other elements have cast doubt on the hypothesis that the Arcturus stream is composed of a homogeneous stellar population (Williams et al. 2009). According to thus more recent work, the putative members do not differ in abundances pattern from the surrounding disk stars. This either indicates a progenitor system that was massive enough to self-enrich to [Fe/H]= -0.6 or, more likely, a dynamical origin for this group (Ramya et al. 2012; Bensby et al. 2014).

To find more MGs in the thick disk and to understand their origins, we are conducting MGs detection using the data of the Large sky Area Multi-Object Fiber Spectroscopic Telescope (LAMOST, so called the Guoshoujing Telescope). LAMOST is a National Major Scientific Project undertaken by the Chinese Academy of Science (Wang et al. 1996; Cui et al. 2012; Zhao et al. 2012). The LAMOST pilot survey conducted from October 2011 to May 2012, obtained several hundred thousand spectra (Luo et al. 2012). Since September 2012, LAMOST has been conducting a general survey, observing about one million stars per year. LAMOST has the capability to observe large, deep and dense regions in the Milky Way Galaxy, which will enable a number of research topics to explore the evolution and the structure of the Milky Way.

In section 2 we describe the data used for MG detection. Section 3  discusses our detection strategy. The analysis of detected MGs is discussed in section 4. A summary of our results is given in section 5.

\section{The Data}


The LAMOST spectra have a resolving power of  R $\sim$ 2000 spanning 3700$\rm \AA \sim $ 9000$\rm\AA$. Two arms of each spectrograph cover this wavelength range with an overlap of 200 $\rm \AA$. The blue spectral coverage is  3700$\rm \AA \sim$ 5900$\rm\AA$ and the red is 5700$\rm \AA \sim$ 9000 $\rm\AA$. The raw data were reduced with LAMOST 2D and 1D pipelines (Luo et al. 2004). These pipelines include bias subtraction, cosmic-ray removal, spectral trace and extraction, flat-fielding, wavelength calibration, sky subtraction, and combination. The radial velocities are measured by cross-correlation between the observed spectra and template spectra from the Elodie library (Moultaka et al. 2004). Stellar parameters (including $T\rm_{eff}$, log $g$, [Fe/H]) are derived by Ulyss \footnote[1]{available at: http://ulyss.univ-lyonl.fr} software package (Wu et al. 2011).  This package enables full spectral fitting for to determine the stellar atmospheric parameters. It minimizes $\chi^{2}$ between an observed spectrum and a template spectrum, and the fit is performed in the pixel space. The method determines all the free parameters in a single fit in order to properly handle the degeneracy between the temperature and the metallicity. 

 Our initial sample was obtained by a cross referencing between  LAMOST DR1 and SDSS DR9. Stars without SDSS photometry and proper motions were eliminated because we need SDSS color to estimate photometric distances. Next, we selected F, G and K main sequence (MS) stars with high signal to noise (S/N $>$ 20) from the initial sample based on color: 0.3 $<$ $(g-r)_{0}\footnote[2]{The subscript nomenclature means dereddened color.}$ $<$ 1.3 and log $g$ $\geqslant$ 3.5.  To eliminate M stars, we added other color constraints, i.e.,  $i-z$ $<$ 0.3 and $r-i$ $<$ 0.53. With the above constraints, 2,09,563 stars were culled. In order to estimate photometric parallaxes of MS stars, we adopted the relation from Ivezi\'{c} et al. (2008), which gives the absolute magnitude in the $r$ band, $M\rm_{r}$ as a function of \textit{g-i} and \textit{[Fe/H]}. Stars with proper motion errors worse than 6 mas/year are also abandoned. The rectangular velocity components relative to the Sun
for these stars were then computed and transformed
into Galactic velocity components \textit{U}, \textit{V}, and \textit{W},
and corrected for the peculiar solar motion
(\textit{U}, \textit{V}, \textit{W}) = (-10.0, 5.2, 7.2) km s$^{-1}$ (Dehnen $\&$ Binney
1998). The UVW-velocity
components are defined as a right-handed system with \textit{U} positive
in the direction radially outward from the Galactic center,
\textit{V} positive in the direction of Galactic rotation, and W
positive perpendicular to the plane of the Galaxy in the
direction of the north Galactic pole. Our typical uncertainties in \textit{U}, \textit{V} and \textit{W} are no
more than $\sim$20 km s$^{-1}$.

Following the Equations (1)-(3) of Bensby et al. (2003) and the characteristic velocity dispersions and
asymmetric drift values given in their Table 1, we  determined
the likelihood that each star belongs to the thin disk, the thick disk and the halo based on its kinematic alone. Ratios of these likelihoods were then
used to find stars that are more likely to be thin disk than thick
disk (i.e., TD/D $<$ 0.50) and more likely to be thick disk and halo than
thin disk (i.e., TD/D $>$ 10).  This strategy yielded a thick disk and halo sample (\textit{d} $<$ 2 kpc) of 7,993 stars.




\section{Moving group detection}
MGs have two main types. One is dynamical streams, i.e., groups of stars that are trapped in a small region of phase space
by dynamical resonances (such as spiral waves or a bar). The stars distribution in such streams in velocity space does not depend on the star's origin, age, or type (Dehnen 1998; Famaey et al. 2005). The other type of MG results from tidal streams, where the stars
originated from the same bound object, such as a cluster or a satellite
galaxy.  But once they become unbound, they will have
slightly different orbits and, hence, frequencies, so that they will
phase-mix. This leads to a broadening of the velocity distribution which is more prominent in the W-component because the vertical
frequency is shorter than the horizontal frequencies.

 One method for identifying MG is to search for overdensities in phase space.  Helmi et al. (1999) and Helmi $\&$ de Zeeuw (2000) presented an  approach involving searching for streams in the space spanned
by the integrals of the motion—that is, the energy E and angular momentum L$_{z}$ vs. L$_{\bot}$ $\equiv$ (L$_{x}^{2}$ + L$_{y}^{2})^{1/2}$. This method is often used to detect halo
streams (Kepley et al. 2007). Dynamic streams tend to be clustered in (\textit{U}, \textit{V}) space (Zhao et al. 2009;  Antoja et al. 2008). Klement et al. (2008) and Arifyanto $\&$ Fuchs (2006) detected streams in
\textit{V} and $\sqrt{U^2+2V^2}$ space. Their strategy for finding nearby stellar streams in velocity space is based on the Keplerian
approximation for orbits, developed by Dekker (1976). In this paper we follow a similar approach.

Fig. 1 presents our detection procedure. The left-top panel plots the distribution of 7,993 stars in \textit{V} and $\sqrt{U^2+2V^2}$ as contours. The bin size is 10 km s$^{-1}$. Color represents the number density in each bin. Several clumps appear in this contour plot. Next, MGs are detected using the wavelet method (Skuljan et al. 1999; Zhao et al. 2009; Klement et al. 2008, 2009). The wavelet transform provides an easily interpretable visual representation of clumps. A mother
wavelet named Mexican hat is the second derivative of a Gaussian and gets good results when
applied to find singularities. The scale of our wavelet transform is adopted as 20 km s$^{-1}$ which is the typical error of velocity measurements.

The top-right panel displays the positive wavelet coefficient. An obvious overdensity locates at \textit{V} $\sim$ -100 km s$^{-1}$ and $\sqrt{U^2+2V^2}$ = 150 km s$^{-1}$, corresponding to the Arcturus group detected by Navarro et al. (2004). There are some other overdensities in this plot.
To test the significance of these, we performed 250 Monte Carlo (MC) simulations using the
same number of stars as in our sample randomly drawn from a Galactic model consisting of  Schwarzschild distributions (Binney $\&$ Tremaine 1987) to represent our sample (see equation 1). The goal was to create a `smooth' reference model velocity distribution that matches the overall velocities of
our sample. To make the distribution of this random sample consistent with our observed sample, we adopted the dispersions ($\sigma_{U}$ and $\sigma_{V}$) and rotational offsets from Local Standard of rest (LSR) as (74, 38, -80) km s$^{-1}$ . Fig. 2 shows a comparison of velocity distribution of our sample (solid line) and the MC sample (dashed line).

\begin{eqnarray}
f \varpropto exp -\frac{1}{2}[(\frac{U}{\sigma U})^2 + (\frac{V-V_{0}}{\sigma V})^2]
\end{eqnarray}

The bottom-left panel of Fig. 1 shows the contour distribution using the wavelet coefficient for a simulated sample. It is clear there are some fake features that arise from the Poission noise or other limits of the method. For each simulated sample, we obtained a wavelet coefficient. Then the average coefficient (wc$_{m}$) of 250 wavelet coefficients and the standard deviations ($\sigma wc_{m}$) are obtained. The final wavelet coefficient (wc$_{f}$) of our real sample was obtained by Equation 2 (wc$_{o}$ represents the coefficient of observed sample; wc$_{m}$ is the mean coefficient of the 250 random sample; $\sigma wc_{m}$ means the standard deviation of the 250 random sample;  $wc_{f}$ is the final coefficient.)

\begin{eqnarray}
 wc_{f}&=&\frac{wc_{o}-wc_{m}}{\sigma wc_{m}}
\end{eqnarray}

The bottom-right panel of Fig. 1 shows the resulting contour distribution of wc$_{f}$. Only those wc$_{f}$ $>$ 2 are ploted. Three groups are apparent and encircled with black lines labeled V1, V2 and V3.  V1 at $\sim$ -100 km $s^{-1}$ is consistent with the Arcturus stream (Navarro et al. 2004).  The range of V1 is from -110 to -70 km s$^{-1}$. The AF06 stream locates at -80 km s$^{-1}$. Tables 1 and 2 in Arifyanto $\&$ Fuchs (2006) show that the velocity and metallicity distributions of the members of the AF06 and Arcturus streams are practically identical. Arifyanto $\&$ Fuchs (2006) think the stars of Arcturus stream and AF06 might originate from the same population based on the colour-magnitude diagrams shown in their Fig. 4. The orbits of those two streams are very similar. This suggests the streams are probably related to each other.  Thus, we regard V1 as a combination of Arcturus stream and AF06. V2, centered at -10 km s$^{-1}$, possibly corresponds to the Hyades-Plaiades stream. V3, centered at -180 km s$^{-1}$,  may be connected with the stream (\textit{V} $\sim$ -160 km s$^{-1}$) found by Klement et al. (2008) using RAVE data (K08 hereafter), however, V3 and K08 are different in L$_{z}$ and L$_{\bot}$. From Fig. 15 in Klement et al. (2008), the center of L$_{\bot}$ is at 850 kpc km $s^{-1}$, while L$_{\bot}$ for V3 ranges [200, 500] kpc km $\rm s^{-1}$ (see Fig. 3). Thus, we regard V3 as a previously undetected stream.

\section{Moving group analysis}


Additional work to understand the formation mechanism of our candidate MGs was undertaken, but hindered by the contamination of field stars. Initial member stars were extracted by range of \textit{V} velocity in bottom-right panel of Fig. 1. Fig. 3 shows a second member stars selection criterion. The left-top panel shows the initial members in L$_{z}$ and L$_{\bot}$. Green plus signs represent V1; blue asterisks represent V2; violet diamonds represent V3. It is clear that the initial members clump in L$_{z}$. However, in the direction of L$_{\bot}$, larger scatter appears in V1 and V3, perhaps caused by contamination of field stars. The other panels in Fig. 3 show histograms of L$_{\bot}$. Generally, MGs should be clustered both in [\textit{V}, $\sqrt{U^2+2V^2}$] and [L$_{z}$, L$_{\bot}$].  The peaks of the L$_{\bot}$ histograms each MG are bracketed by dashed lines. Thus, the final members of each MG are those within the two dashed lines: V1 [100, 500] kpc km s$^{-1}$; V2 [600, 800] kpc km s$^{-1}$; V3 [200, 500] kpc km s$^{-1}$.

Fig. 4 presents these three groups in a different way. The symbols have the same meaning as those of Fig. 3. The small black dots represent the non-MG stars in the whole sample. 
The top panel provides their distribution in [Fe/H] and \textit{W}. Most stars in V3 are metal poor; [Fe/H] $<$ -0.5 dex. The \textit{W} velocity distributions in V1 and V3 show larger dispersions than V2. The middle panel presents the distribution in [\textit{U}, \textit{V}]. V1 and V3 show very broad dispersions in U while V2 has a small dispersion perhaps because the stars in V2 have been perturbed by the Galaxy's bar. The bottom of Fig. 4 plots the [Fe/H] and vertical distance of these MGs. The distance of most stars in our sample is larger than 200 pc. However, stars in V2 are relatively nearby. All these MGs extend to about 1.5 kpc.

Fig. 5 presents the eccentricity distribution of three candidate streams. Galactic orbits were calculated with the Milky Way potential model of Allen $\&$ Santillan (1991). The model
is time-independent, axisymmetric, fully analytic, and consists
of a spherical central bulge, a disk, a massive spherical halo,
and has a total mass of 9$\times$10$^{11}$ solar masses. Output parameters are: the minimum and maximum distances from the Galactic center;
R$_{min}$ and R$_{max}$ (i.e., the peri- and apocenter); the maximum distance from the Galactic plane Z$_{max}$
and the eccentricity, e=(R$_{max}$-R$_{min}$)/(R$_{max}$+R$_{min}$).

The first three panels in Fig. 5 show the distribution of the total sample in  eccentricity vs. L$_{z}$ and in eccentricity vs. [Fe/H]. The eccentricity and metallicity of our sample has a very wide dispersion. The last three panels present the normalized histogram of our three streams, respectively. The black solid lines represent the total sample, while the green dashed lines denote our candidate MGs. As expected, the MGs have very narrow distribution of eccentricity. The peak of V2 is e$\sim$ 0.1, suggesting this stream is not debris of an accretion event. The peak of V3 is e $\sim$ 0.8, as would be expected structure resulting from an accrete event. The peak of V2 is e $\sim$ 0.4, similar to the peak of the total sample. Generally dynamical instabilities in the Galactic disc have been found to involve velocities in the range \textit{U} and \textit{V}$\sim \pm$  50 km s$^{-1}$. At this time there is no evidence that spiral or bar structure can cause high-velocity streams like Arcturus-AF06. This is why they are usually attributed to merger events. The Arcturus stream has been interpreted as originating from the debris of a disrupted satellite (Navarro, Helmi $\&$ Freeman 2004; Helmi et al. 2006). AF06 was assigned similar origin, based on its kinematics.  However, a detailed investigation by Williams et al. (2009) found that the chemical abundances are consistent with a dynamical origin but do not entirely rule out a merger
one. We conclude V1 and V3 have different origins than V2. In the L$_{z}$ and  L$_{\bot}$ distributions for the stars extracted by \textit{V} velocity (Fig. 3),  V2 shows a small dispersion, while V1 and V3 have very broad dispersions. Moreover, the \textit{U} component of V1 spans [-100, +100], which could not be induced by a bar. Thus, we tentatively regard V1 and V3 as products of accretion event.

\section{Conclusion}
Three candidate moving groups were detected with high confidence from 7,993 thick disk and halo sample using LAMOST DR1 data. The velocity distribution of the sample spans distances to $\sim$ 2 kpc from the sun.  Using a wavelet technique, three significant local kinematic groups are detected. Two of them correspond to the Hyades-Pleiades and Arcturus-AF06 streams. The other is a possible new stream centered at \textit{V} $\sim$ -180 km s$^{-1}$. Among these three MGs, the Arcturus-AF06 stream is the most prominent and dominates the whole sample. The MGs show different W components and very narrow eccentricity distributions.  We tentatively concluded that the Hyades-Pleiades has a dynamical origin, perhaps the result of  perturbation by Galaxy Bar or spiral arms.  The new stream is metal poor and has a high eccentricity, suggesting a debris of a satellite accetion event. Although the origin of Arcturus-AF06 is not very definitely determined, we provided some evidence that this stream may also be a debris form a satellite accretion event.

\acknowledgments

This study  is supported by the
National Natural Science Foundation of China under grant No. 11390371, 11233004, 11222326, 11150110135, 11103034 and the National Key Basic Research Program of China (973 program) 2014CB845701 and 2014CB845703. Support from the US National Science Foundation (AST-1358787) to Embry-Riddle Aeronautical University is acknowledged. Guoshoujing Telescope (the Large Sky Area Multi-Object Fiber Spectroscopic Telescope LAMOST) is a National Major Scientific Project built by the Chinese Academy of Sciences. Funding for the project has been provided by the National Development and Reform Commission. LAMOST is operated and managed by the National Astronomical Observatories, Chinese Academy of Sciences.

\clearpage



%


\begin{figure}
\plotone{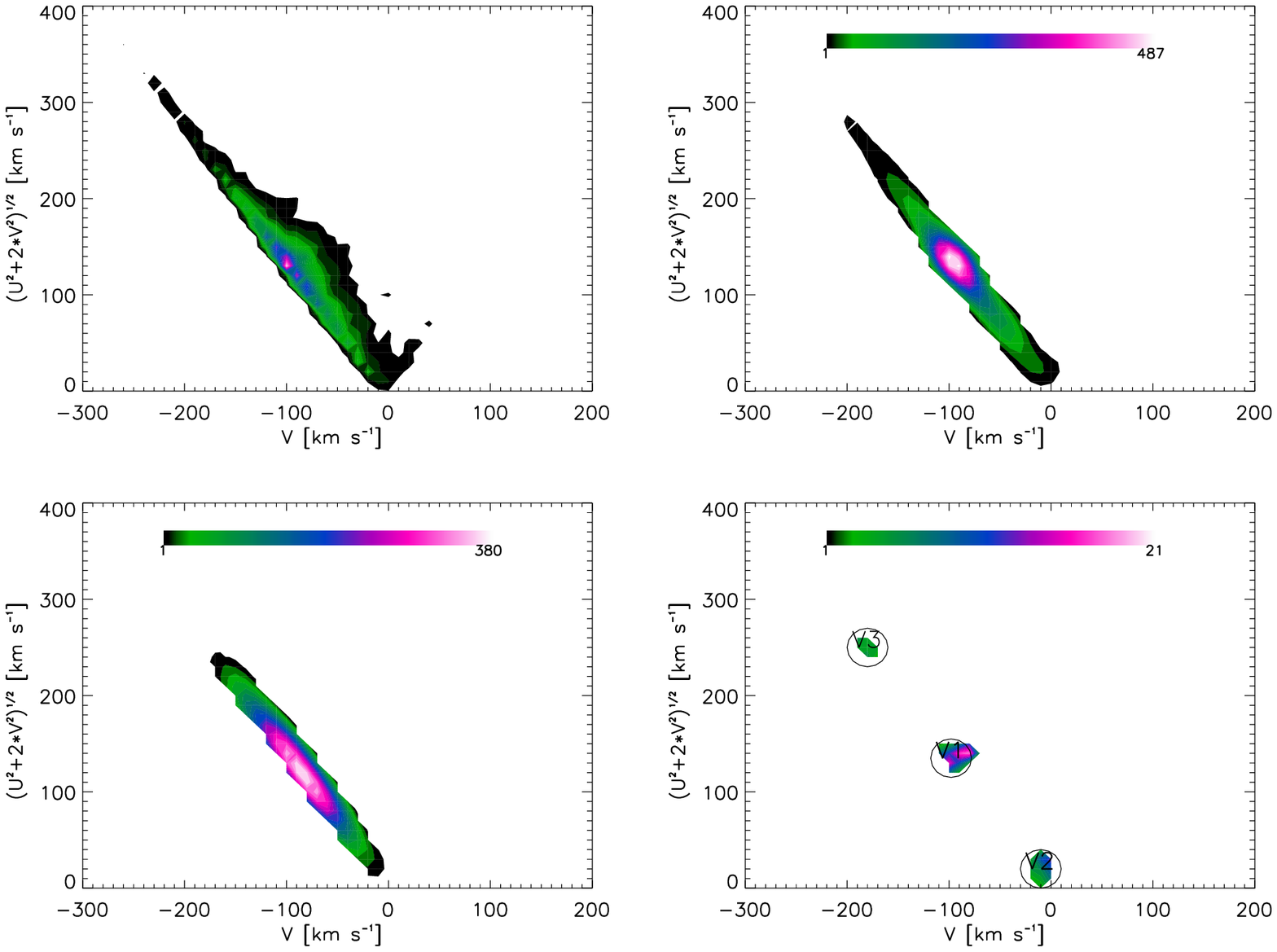}
\caption{Moving groups detection using thick disk and halo sample. Top-left: contour plot of our sample in \textit{V}  and $\sqrt{U^2+2V^2}$. Top-right: positive wavelet maps in the system of (\textit{V}, $\sqrt{U^2+2V^2}$) coordinates for our sample. Bottom-left: positive waveletmaps in the system of (\textit{V}, $\sqrt{U^2+2V^2}$) coordinates for MC sample. Bottom-right: significance of the overdensities seen in top-right panel, obtained as described in the text. Note that only areas with $\sigma$ $>$  2 are displayed. We encircled in black
and labeled all features (V1, V2 and V3) that we consider to be stellar streams.    \label{fig2}}
\end{figure}
\clearpage

\begin{figure}
\plotone{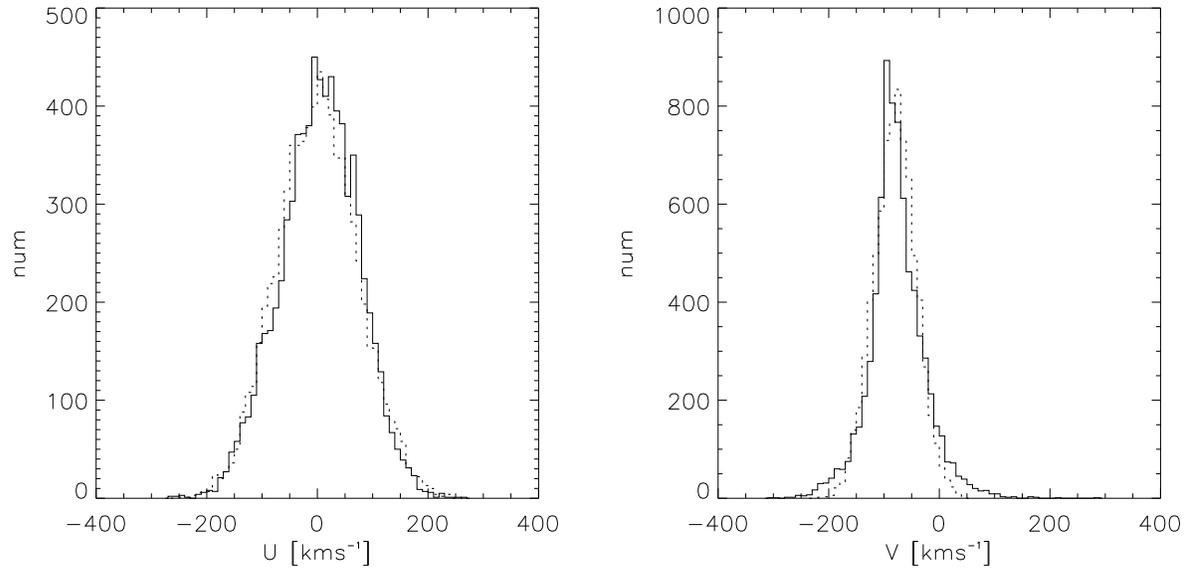}
\caption{Velocity distributions of our sample compared with one MC sample. Solid lines represent the distributions of observed sample while dashed lines represent the distribution of MC sample.  The MC sample consists of Schwarzschild distributions for the thick disks and the halo.    \label{fig2}}
\end{figure}
\clearpage

\begin{figure}
\plotone{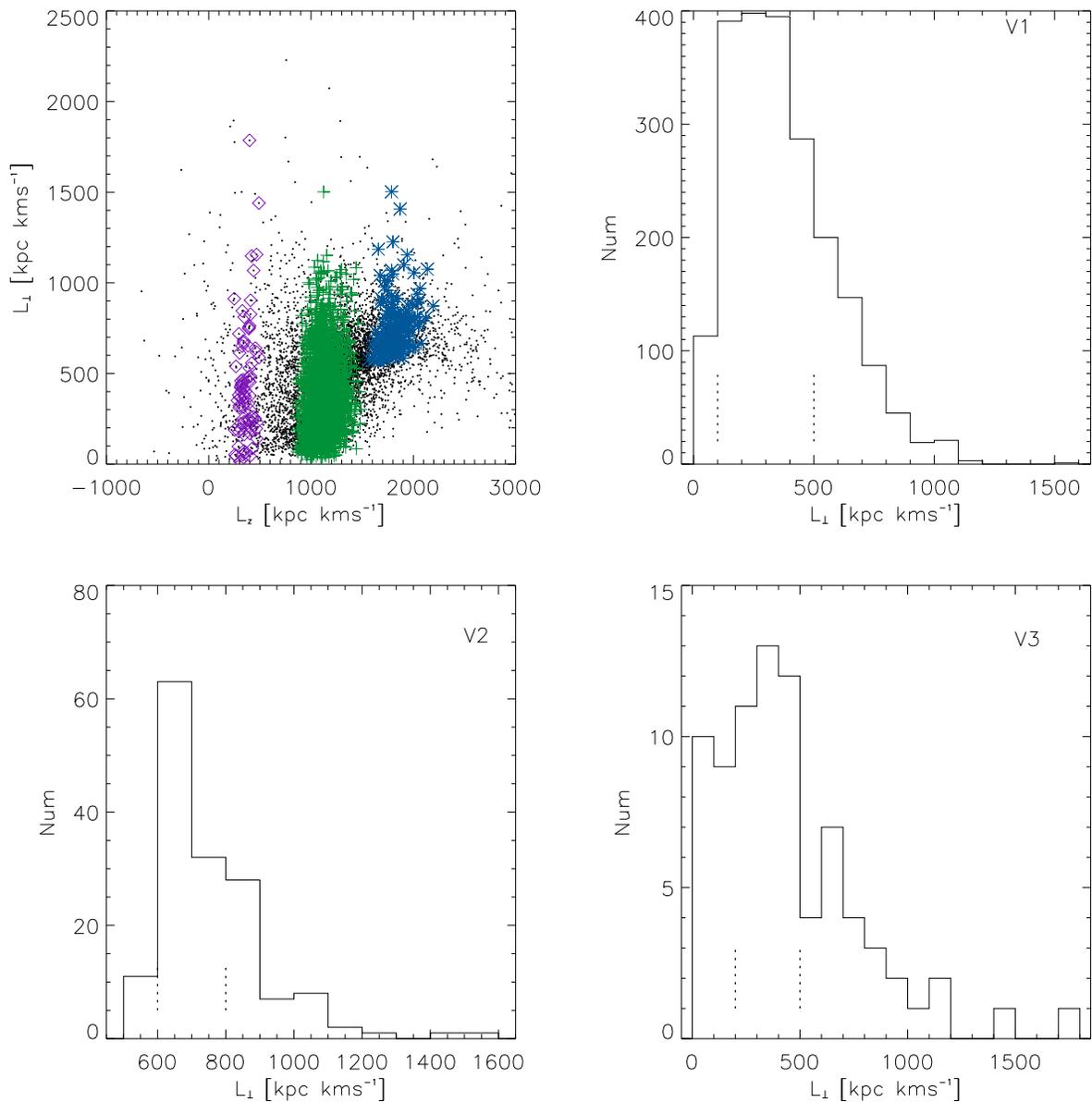}
\caption{The L$_{\bot}$ distribution of the MGs. Top-left: L$_{\bot}$ distribution for initial member stars extracted only by \textit{V} velocity. Green plus sign represents V1; blue asterisk represents V2; violet diamond represents V3; small black dots represents the stars of the whole sample. Other panels give the histograms of three MGs. Dashed lines indicate the peaks of the L$_{\bot}$ distribution for these three MGs.\label{fig2}}
\end{figure}

\clearpage

\begin{figure}
\plotone{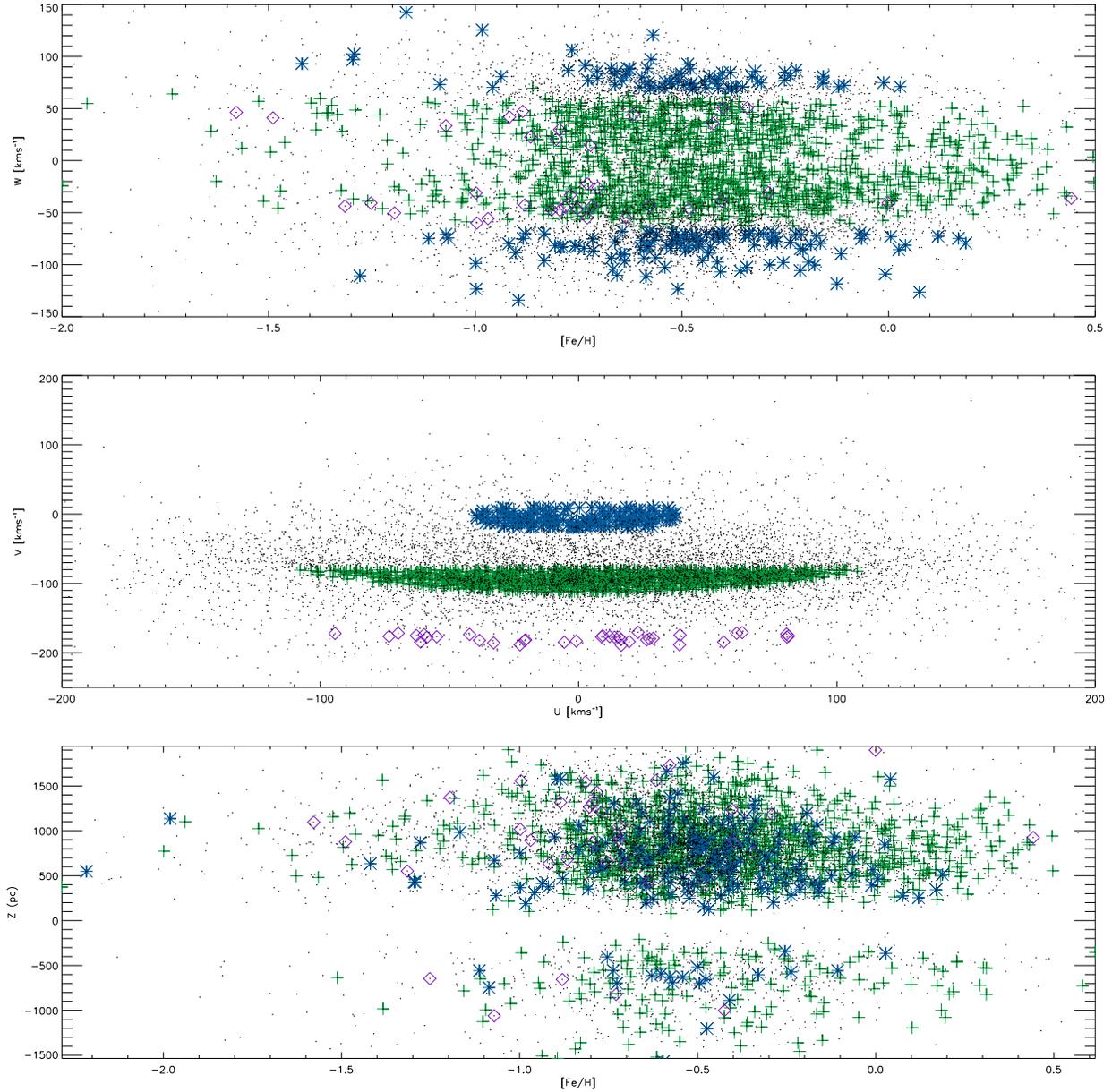}
\caption{Characters of three streams. The symbols have the same meanings with those in Fig. 2.  Top: distribution of three streams in [Fe/H] and W.
Middle: distribution of three streams in [\textit{U}, \textit{V}]. Bottom: distribution of three streams in [Fe/H] and Z  .\label{fig2}}
\end{figure}
\clearpage

\begin{figure}
\plotone{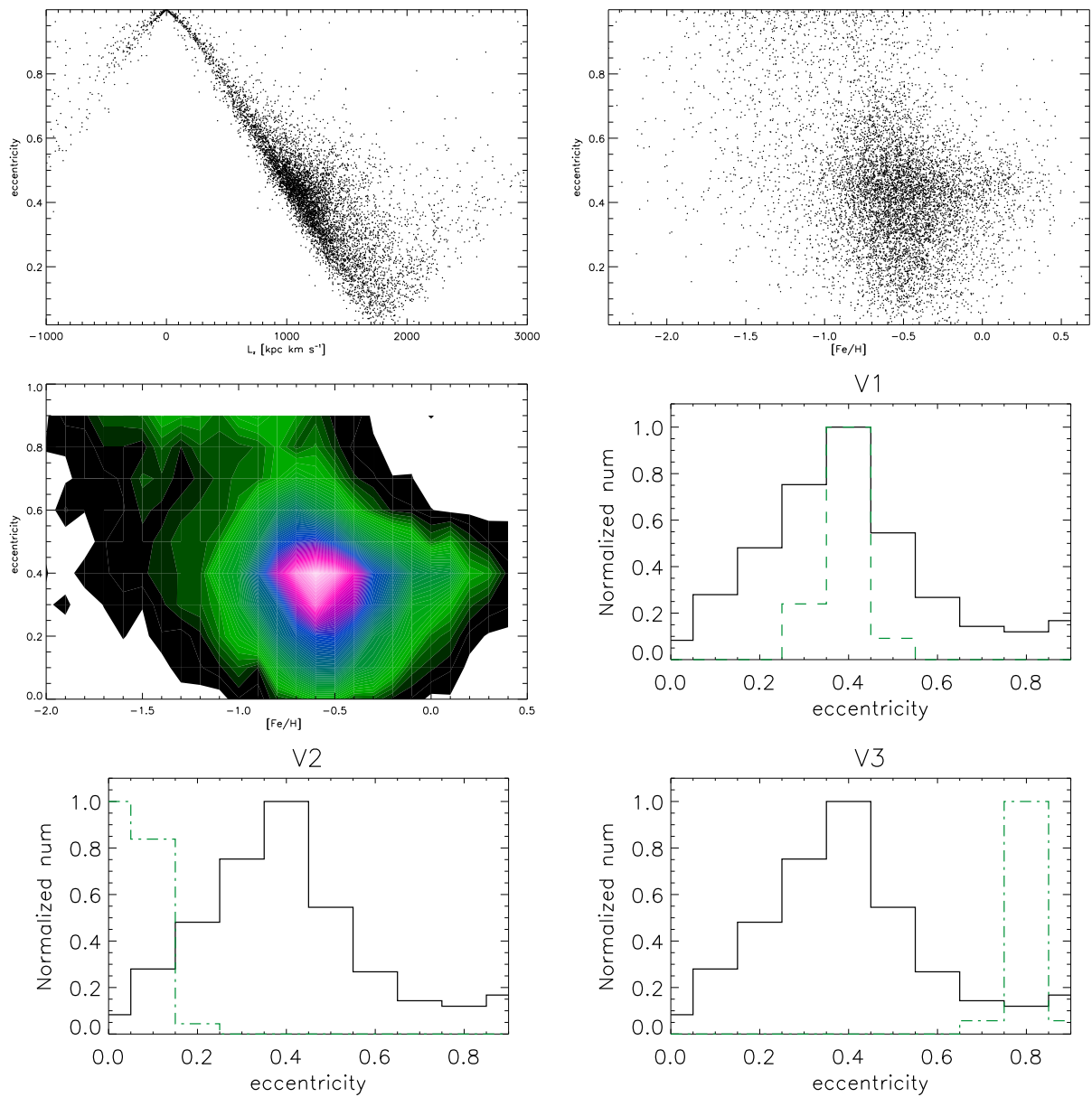}
\caption{Eccentricity distribution of three streams. The first three panels show the distribution of total sample in L$_{z}$ vs. eccentricity; in [Fe/H] vs. eccentricity. The black solid line in the last three panels is the histogram of the total sample while green dashed line presents the histogram of three streams, respectively.\label{fig2}}
\end{figure}
\clearpage








\clearpage





\end{CJK}
\end{document}